# When Time-Reversible Sensor Records Require Hidden Dissipation

Andrey V. Dobrynin*[1] and Anastasia Stroujkova[2]

[1] *Department of Chemistry, University of North Carolina, Chapel Hill, NC, 27599, United States*

[2] *SeisLogic, LLC, Chapel Hill, NC, 27517, United States*

email: avd@email.unc.edu

**Abstract:** A binary sensor record can be time reversible yet require dissipation in every compatible finite-state Markov model. For a four-state sensor with correlated dwell times, two observed switching statistics give an exact equilibrium criterion and, when equilibrium is excluded, a positive lower bound on entropy production independent of hidden-state number. An explicit four-state construction provides an upper bound on the minimum dissipation. For a noisy Poisson sensor, the bounds differ by less than a factor of two and vanish at a kinetic matching point while the sensor remains dissipative.

*Introduction.*— Physical sensors generally observe only a fraction of the microscopic transitions occurring in their environment. Ion channels, molecular detectors, and electronic switches can produce binary records while their hidden degrees of freedom sustain probability currents and dissipate energy. Time-reversal asymmetry of observed trajectories can reveal irreversible dynamics [1-4], and waiting-time statistics provide additional information when visible currents vanish [5]. However, a time-reversible observed record does not necessarily allow an equilibrium description of the underlying dynamics.

Different microscopic processes with hidden states can generate identical observable trajectories, including waiting-time distributions and correlations [6-11]. The dissipation of a particular microscopic process must therefore be distinguished from the minimum dissipation required by the observed trajectory. If an equilibrium process reproduces the complete observed record, hidden dissipation cannot be established from that record alone. If no such process exists, the observations establish that hidden dissipation is necessary even when the measured trajectory is time reversible. This raises the question of whether a complete stationary binary record can be reproduced by a finite-state equilibrium Markov process[12-14].

Here we consider a sign-symmetric four-state Markov model consisting of coupled binary hidden and observed variables. The complete observed record is time reversible for all positive transition rates. However, this does not guarantee an equilibrium description. We show that three observable kinetic quantities determine the complete recorder path law, while only two decide equilibrium realizability. When equilibrium is excluded, short-time persistence provides a positive lower bound on entropy production for all compatible finite-state Markov models. We also construct the four-state model with the lowest entropy production, which establishes an upper bound on the dissipation required to reproduce the observed record.

For a noisy Poisson sensor, these bounds differ by less than a factor of two and vanish linearly at a kinetic matching point.

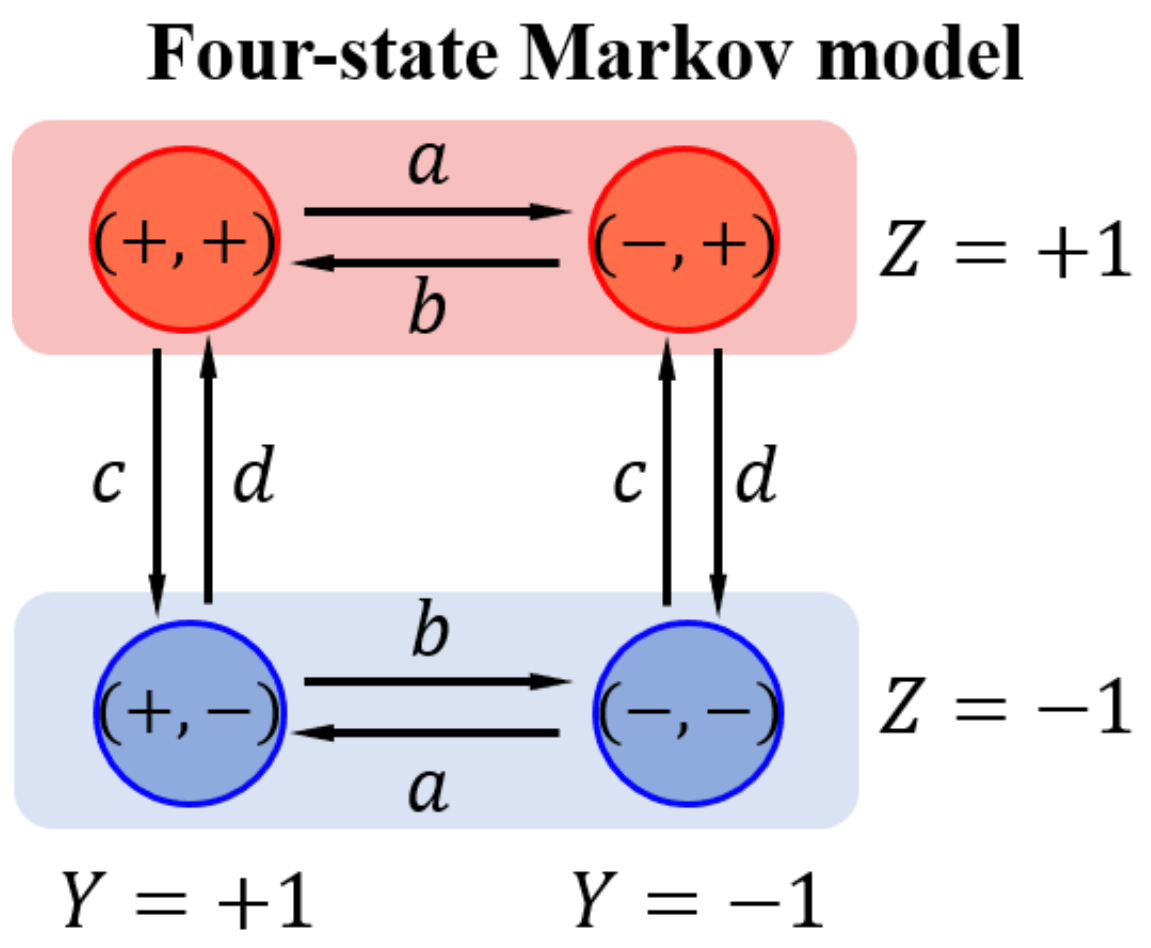

FIG. 1. Four-state Markov model for a hidden variable $Y = \pm 1$ and an observed variable $Z = \pm 1$. The four nodes are labeled by the pair $(Y, Z)$. Horizontal transitions correspond to flips of the hidden variable $Y$ while vertical transitions correspond to flips of the observed variable $Z$. The rates $a$ and $c$ describe hidden and observed flips from aligned states $Y = Z$, respectively. The rates $b$ and $d$ describe corresponding flips from misaligned states $Y \neq Z$. For the noisy Poisson sensor, the corresponding rates are $a = b = \gamma$, $c = f(1-\eta)/2$, and $d = f(1+\eta)/2$. For $\mu < s$, an equilibrium process reproducing the same observed trajectory is obtained by replacing the transition rates with their equilibrium values given in Eq. (9).

*Model.*—Consider a stationary continuous-time Markov process described by two binary variables $Y = \pm 1$ and $Z = \pm 1$, representing the hidden state and the recorded signal, respectively (Fig. 1). We allow only transitions in which one variable changes at a time and the model is symmetric under the transformation $(Y, Z) \to (-Y, -Z)$. The resulting four-state model is characterized by four positive transition rates. The hidden variable flips at rate $a$ when $Y = Z$ and at rate $b$ when $Y \neq Z$, while the observed variable flips at rates $c$ and $d$ in the corresponding states (Fig. 1). These transition rates couple the hidden and observed dynamics without necessarily satisfying the detailed balance condition.

The observed trajectory consists of intervals with fixed $Z$, separated by transitions at which $Z$ changes sign. During each interval, the hidden alignment evolves according to the generator $T$, while a visible flip changes the alignment according to the matrix $M$. In the aligned and misaligned basis, these matrices are

$$T = \begin{pmatrix} -a-c & a \\ b & -b-d \end{pmatrix}, \qquad M = \begin{pmatrix} 0 & c \\ d & 0 \end{pmatrix}. \tag{1}$$

The diagonal terms of $T$ account for visible flips that determine a dwell interval. The matrix $M$ describes transitions between successive dwell intervals, while $e^{Tt}$ describes the evolution of the hidden state in the absence of visible flips.

We introduce the four combinations of transition rates:

$$S = a + b + c + d, \quad s = \sqrt{cd}, \quad \mu = (ad + bc + 2cd)/S, \;\; \pi = (b + d, a + c)/S. \tag{2}$$

The parameter $\mu$ gives the mean rate of visible flips in the stationary state. The stationary distribution of hidden alignment, conditioned on either value of $Z$, is $\pi$. For an observed trajectory with successive dwell intervals $t_0, t_1, \ldots, t_n$ including the first and last intervals truncated by the observation window, the probability density is proportional to

$$\mathcal{L}(t_0, \ldots, t_n) = \pi e^{Tt_0} M e^{Tt_1} M \cdots M e^{Tt_n} \mathbf{1}, \tag{3}$$

where $\mathbf{1}$ denotes the column vector with both components equal to unity. This matrix product retains the correlations between successive dwell intervals without requiring a renewal approximation.

The microscopic dynamics contain four independent rates, but the complete stationary observed trajectory depends only on $S$, $s$, and $\mu$. These three quantities can be determined from the observed kinetics. In particular, $s^2$ is determined by the curvature of the stationary persistence probability at zero time. Away from the kinetic matching condition introduced below, the third derivative of the persistence probability determines $S$ and hence the two decay rates of the matrix $T$. Thus, the complete observed trajectory is characterized by three quantities without direct measurement of the hidden variable $Y$.

Different combinations of transition rates $a$, $b$, $c$, and $d$ can produce the same $S$, $s$, and $\mu$, resulting in identical observed trajectory statistics. Within the four-state model, the complete dwell-time distribution is sufficient to determine the statistics of the observed trajectory, even when successive intervals are correlated. For more general hidden-state models, however, a few dwell-time moments, the single-dwell distribution, or a two-time correlation function cannot establish the statistics of the complete trajectory[15].

*Equilibrium realizability.*— We first determine the conditions under which the observed trajectory can be reproduced by a finite-state equilibrium Markov process. The complete observed trajectory is time reversible for all positive values of $a$, $b$, $c$, and $d$. However, this does not imply that the four-state model satisfies detailed balance. The existence of an equilibrium process reproducing the observed trajectory depends on the relation between $\mu$ and $s$. We find

| Condition | Visible process | Equilibrium realization |
|---|---|---|
| $\mu > s$ | non-Markovian | no finite-state equilibrium realization |
| $\mu = s$ | telegraph process, rate $s$ | two states |
| $\mu < s$ | non-Markovian | explicit four-state realization |

The classification applies to the complete stationary observed trajectory, including correlations between successive dwell intervals. For $\mu \neq s$, persistence statistics determine all three quantities $S$, $s$, and $\mu$. At $\mu = s$, the observed signal reduces to a symmetric telegraph process, and $S$ cannot be determined from the observations [15].

The stationary persistence probability $g(t)$ is defined as the probability that the observed state does not change during a time interval $t$. For the four-state model, we obtain [15]

$$g(t) = \exp(-ht)\left[\cosh(\omega t) + \frac{(h-\mu)}{\omega}\sinh(\omega t)\right], \tag{4}$$

where $h = S/2$ and $\omega^2 = h^2 - S\mu + s^2 > 0$. For a finite-state equilibrium Markov process with observed variables even under time reversal, $g(t)$ is a sum of exponential relaxation modes with nonnegative weights. This requires $g(2t) - g(t)^2 \geq 0$, independently of the number of hidden states, as derived in the Supplemental Material [15]:

$$g(2t) - g(t)^2 = \frac{(s^2 - \mu^2)}{\omega^2}\exp(-2ht)\sinh^2(\omega t). \tag{5}$$

Equation (5) shows that the sign of the persistence covariance is determined by $s^2 - \mu^2$. For $\mu < s$, the covariance is positive, consistent with equilibrium relaxation. For $\mu > s$, the covariance is negative for $t > 0$, ruling out an equilibrium description by any finite-state Markov model with observed variables even under time reversal. Thus, the observed trajectory can be time reversible without allowing an equilibrium description.

The physical origin of this result is the balance between visible transitions and their reverse transitions. In equilibrium, reverse transitions contribute to the curvature of the persistence probability, requiring $g''(0) \geq [g'(0)]^2$. Since $-g'(0) = \mu$ and $g''(0) = s^2$, equilibrium requires $s \geq \mu$. For $\mu > s$, the observed signal has fewer short time return events than an equilibrium process with the same mean switching rate. This provides a direct signature of hidden dissipation without observing the hidden transitions.

*Thermodynamic cost of hidden dynamics.*—The interval $\mu > s$ permits a stronger conclusion than the absence of an equilibrium realization. The observed short-time kinetics impose a positive lower bound on entropy production for every compatible finite-state Markov process. We characterize the observed trajectory by the stationary probability distribution $P$. The minimum possible entropy production rate, $\sigma_{\text{inf}}(P)$, is defined as the infimum over all stationary finite-state Markov processes whose binary observations exactly reproduce $P$. The microscopic and observed variables are taken to be even under time reversal. For a Markov process with transition rate matrix $Q = (q_{ij})$ and stationary probabilities $p = (p_i)$, the entropy production rate is

$$\sigma(Q,p) = k_B \sum_{i<j} \left(p_i q_{ij} - p_j q_{ji}\right) \ln \left[\frac{p_i q_{ij}}{p_j q_{ji}}\right]. \tag{6}$$

Unoccupied states are excluded, transitions between states may be forbidden, and irreversible transitions carrying a finite stationary flux produce infinite entropy production. These conditions define the admissible finite-state Markov processes without imposing a network topology.

The persistence derivatives give the bound

$$\sigma_{\text{inf}}(P) \geq \underline{\sigma} = 2k_B\delta \ \text{artanh}\left(\frac{\delta}{\mu}\right) > 0, \qquad \delta = \sqrt{\mu^2 - s^2}, \qquad \mu > s. \tag{7}$$

Equation (7) is independent of the number of microscopic states and does not require a maximum transition rate. The bound follows from the transition fluxes between observed states, which are constrained by $\mu$ and $s^2$, and the convexity of the entropy production cost [15].

For $\mu \leq s$, the explicit equilibrium constructions below show that zero-dissipation realizations exist. In the case $\mu > s$, the deficit of immediate returns requires uncompensated microscopic fluxes and hence positive entropy production. As the observed kinetics approach $\mu = s$ from above, the lower bound continuously decreases to zero, even though the underlying dynamics need not approach detailed balance.

The bound $\underline{\sigma}$ is optimal based on the short-time persistence statistics $\mu$ and $s$. However, no compatible finite-state process attains this bound for $\mu > s$. Four-state models with fixed $\mu$ and $s$ approach it as $S \to \infty$ [15]. The complete observed record contains additional information through $S$, which may impose a stronger restriction on entropy production. Related bounds based on visible transitions and waiting-time statistics are given in Refs. [2, 16, 17].

The absence of a finite-state equilibrium realization does not exclude the possibility of arbitrarily small dissipation as the number of hidden states increases. Equation (7) rules out this possibility whenever $\mu > s$, establishing a strictly positive lower bound $\sigma_{\text{inf}}(P) \geq \underline{\sigma} > 0$ independent of hidden-state number. Thus, increasing the complexity of hidden dynamics cannot reduce dissipation below this bound, even when the observed trajectory is time reversible. In the Supplemental Material [15], we apply this bound to an alternating renewal process, establishing a positive dissipation infimum over all hidden-state dimensions.

For $\mu < s$, the complete observed trajectory can also be reproduced by a process satisfying detailed balance. To demonstrate this, we introduce

$$k = \frac{ad + bc}{2s}, \qquad q = h\sqrt{1 - \frac{\mu^2}{s^2}}, \qquad \rho = \frac{h - q}{k + s} > 0. \tag{8}$$

The corresponding equilibrium transition rates are

$$a_{\text{eq}} = k\rho, \qquad b_{\text{eq}} = \frac{k}{\rho}, \qquad c_{\text{eq}} = s\rho, \qquad d_{\text{eq}} = \frac{s}{\rho}. \tag{9}$$

These rates satisfy detailed balance and reproduce the original values of $S$, $s$, and $\mu$. Thus, both processes generate identical stationary observed trajectories, including correlations between successive dwell times, despite having different microscopic transition rates and interpretations of the hidden variable. At $\mu = s$, a two-state equilibrium process reproduces the symmetric telegraph trajectory with switching rate $s$. Therefore, time reversibility of the observed trajectory does not establish equilibrium of the actual sensor, since the same observations can arise from equilibrium and nonequilibrium dynamics.

To examine the additional constraint imposed by $S$, we minimize entropy production over all sign-symmetric four-rate models with fixed $S$, $s$, and $\mu$. These models reproduce the same observed trajectory statistics but may have different microscopic currents and entropy production rates. In the interval $\mu > s$, using the definitions of $h$ and $\delta$, the transition rates

$$(a_0, b_0, c_0, d_0) = (h - \mu + \delta,\ h - \mu - \delta,\ \mu - \delta,\ \mu + \delta) \tag{10}$$

minimize entropy production within this class of models. In this construction, the hidden flip rates may depend on alignment, allowing the observed state to influence the dynamics of the hidden variable.

The exact minimum entropy production over sign-symmetric four-rate models reproducing the same observed record is [15]

$$U = \sigma_{\min}^{(4)} = k_B \delta \ln\left(\frac{K + S\delta}{K - S\delta}\right), \qquad K = ad + bc = S\mu - 2s^2. \tag{11}$$

Since $S$, $s$, and $\mu$ uniquely determine the complete observed record away from the Markovian matching condition, $U$ is the minimum entropy production among all equivalent sign-symmetric four-rate models. Additional hidden states may further reduce entropy production, but the minimum over all finite-state Markov processes remains unknown. Combining Eq. (11) with the lower bound in Eq. (7), we obtain

$$\underline{\sigma} \leq \sigma_{\text{inf}}(P) \leq U, \qquad \mu > s. \tag{12}$$

*Noisy sensor and kinetic matching.*—We now apply this framework to a sensor monitoring an independently switching environment. The environment $Y = \pm 1$ follows a symmetric telegraph process with switching rate $\gamma$. The sensor measures the environment at Poisson-distributed times with frequency $f$. Each measurement agrees with the current environmental state with probability $(1 + \eta)/2$, where $0 < \eta < 1$. Between measurements, the sensor retains its last reading $Z$. Measurements that do not change the reading are invisible in the observed trajectory. The model belongs to the four-state family with

$$a = b = \gamma, c = f(1 - \eta)/2, d = f(1 + \eta)/2. \tag{13}$$

The environmental switching rates are independent of the sensor reading, while the sensor switching rates depend on its alignment with the environment. The physical sensor dissipates energy for all positive $f$ and $\eta$. Its entropy production rate and the kinetic matching frequency are [15]

$$\sigma_{\text{sensor}} = \frac{2k_B \gamma f}{2\gamma + f} \eta \text{ artanh}(\eta) > 0, \quad f_0 = \frac{2\gamma}{\sqrt{1 - \eta^2}}. \tag{14}$$

Substitution into Eq. (2) gives $\mu > s$ for $f < f_0$. Thus, below $f_0$, the observed trajectory cannot be reproduced by any finite-state equilibrium Markov process. For $f > f_0$, the equilibrium construction in Eq. (9) reproduces the complete stationary observed dynamics. Increasing the sampling frequency through $f_0$ changes whether an equilibrium description is possible, while the physical sensor continues to dissipate energy.

In the interval $f < f_0$, the four-rate minimum of Eq. (11) provides an upper bound on the minimum compatible entropy production. Defining

$$\chi = \eta\sqrt{1-\left(\frac{f}{f_0}\right)^2}\,, U(f) = \frac{2k_B\gamma f}{2\gamma + f}\chi\,\mathrm{artanh}(\chi), \tag{15}$$

we obtain

$$\underline{\sigma} \leq \sigma_{\mathrm{inf}}\big(P_f\big) \leq U(f), \tag{16}$$

where $P_f$ describes the stationary observed trajectory at sampling frequency $f$. As $f$ approaches $f_0$ from below, $U(f)$ decreases continuously to zero, while the physical sensor maintains a positive entropy-production rate, $\sigma_{\mathrm{sensor}}(f_0) > 0$. For $f \geq f_0$, an equilibrium description becomes possible and $\sigma_{\mathrm{inf}}\big(P_f\big) = 0$.

For the frequency range $0 < f < f_0$, the upper bound satisfies $U(f) < 2\sigma_{\mathrm{inf}}\big(P_f\big)$, restricting the minimum compatible entropy production to within a factor of two. As $f$ approaches $f_0$ from below, the minimum entropy production vanishes linearly with $(f_0 - f)$. With $S_0 = 2\gamma + f_0$, The lower and upper bounds vanish linearly, with slopes $4k_B\eta^2\gamma f_0/S_0^2$ and $4k_B\eta^2\gamma/S_0$ , respectively [15].

For $\eta = 1/2$ and $f = \gamma$, the physical sensor dissipates at a rate of $0.183k_B\gamma$, compared with the four-rate minimum $U = 0.146\ k_B\gamma$ and the universal lower bound $\underline{\sigma} = 0.102k_B\gamma$. Thus, the unrestricted minimum lies within a factor of approximately 1.43. In the limit of rare sampling, the bounds converge to [15]

$$\sigma_{\mathrm{inf}}\big(P_f\big) \sim k_B f\eta\,\mathrm{artanh}(\eta)\,, \qquad f/\gamma \to 0. \tag{17}$$

This agrees with the leading-order entropy production of the physical sensor, which is therefore optimal to leading order as sampling becomes rare, even though its finite-frequency dissipation can be higher than that of an equivalent microscopic process.

At the matching frequency $f = f_0$, the observed signal becomes a symmetric telegraph process with switching rate $\gamma$, identical to that of the environment [18, 19]. However, the sensor does not measure the environment perfectly. For $\eta = 1/2$, the probability of incorrect reading remains approximately 36.6%. At this frequency, the complete history of sensor readings provides no additional information about the current environmental state beyond the latest reading [15]. Thus, the current reading is sufficient for estimating the environment, even though measurement errors remain. Kinetic matching eliminates the additional information contained in earlier readings, not the measurement noise.

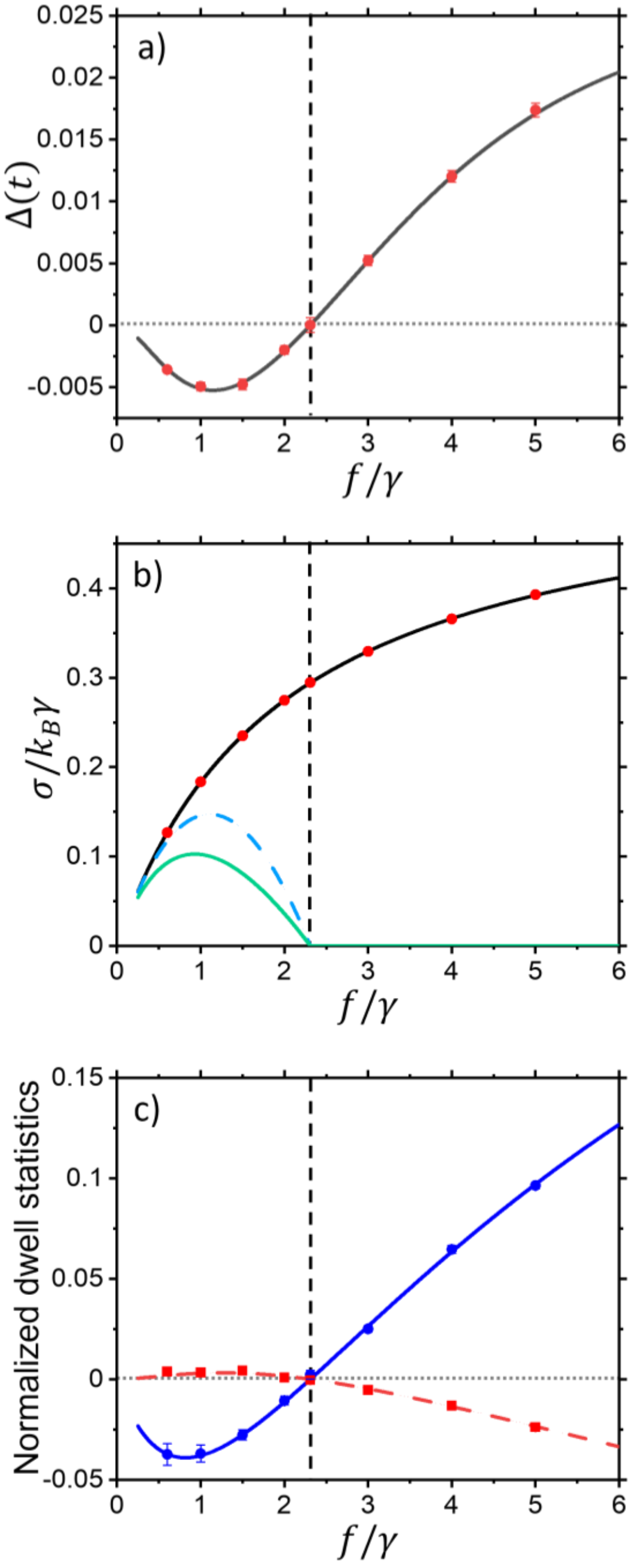

FIG. 2. Kinetic boundary and dissipation bounds for $\eta = 1/2$. Solid and dashed curves show exact results; symbols represent Gillespie estimates from 15 independent stationary trajectories per sampling rate, each of duration $3 \times 10^5/\gamma$. Symbols show corresponding Monte Carlo estimates, and error bars indicate 95% confidence intervals. Vertical dashed lines indicate $f_0/\gamma = 2/\sqrt{(1-\eta^2)}$. (a) $\Delta(t)$ for $\gamma t = 0.5$, where $\Delta(t) = C(2t) - C(t)^2$. (b) Entropy-production rate of the actual sensor (black line), minimum $U$ among observationally equivalent sign-symmetric four-rate models (blue dashed line), and architecture-independent lower bound $\underline{\sigma}$ (green solid line). (c) Excess complete-dwell dispersion (blue) and normalized adjacent dwell covariance (red).

*Numerical tests and observable signatures.*— Figure 2 illustrates the kinetic classification for a noisy sensor with ($\eta = 1/2$). In addition to the persistence probability $g(t)$, we examine the stationary sensor autocorrelation $C(t) = \langle Z(0)Z(t)\rangle$ and the quantity $\Delta(t) = C(2t) - C(t)^2$. The sign of $\Delta(t)$ changes at the matching frequency $f_0$, providing an observable signature of the equilibrium restriction.

Figure 2(a) compares the exact $\Delta(t)$ with estimates from stochastic trajectories at $\gamma t = 0.5$. The sign changes at $f_0$, while the physical sensor continues to dissipate energy. Figure 2(b) compares the sensor entropy production with the upper bound $U$ and the universal lower bound $\underline{\sigma}$. For interval $f < f_0$, these bounds constrain the minimum compatible entropy production and both vanish as $f$ approaches $f_0$, while the sensor dissipation remains positive. For $f \geq f_0$, the minimum compatible entropy production is zero.

The kinetic boundary also appears in the statistics of visible dwell times. Figure 2(c) shows that the excess complete-dwell dispersion and normalized covariance between successive dwell times change sign at $f_0$. These correlations are important because the observed process is generally not a renewal process. The numerical results agree with Gillespie simulations of 15 independent stationary trajectories, each of duration $3 \times 10^5/\gamma$, at each sampling frequency. Error bars indicate 95% confidence intervals. The simulations confirm the predicted kinetic behavior, while the exact equivalence of the observed trajectories is established analytically.

*Discussion.*— Our results demonstrate that time reversibility of an observed trajectory does not establish equilibrium of the underlying dynamics. For interval $\mu > s$, the short time persistence statistics exclude an equilibrium description and impose a positive lower bound on entropy production. For $\mu \leq s$, an equilibrium process can reproduce the complete observed trajectory, including correlations between successive dwell times.

These findings complement the work of Skinner and Dunkel, who derived entropy production bounds from dwell-time fluctuations by optimizing over hidden networks [2, 16]. We extend this perspective by establishing exact conditions under which the complete observed trajectory of a four-rate model with correlated dwell times can be reproduced by an equilibrium process. When equilibrium is excluded, short time persistence provides an analytic lower bound on entropy production, independent of the number of hidden states and their transition rates. This bound is optimal for the two measured quantities, while an explicit four-state construction provides an upper bound on the minimum required dissipation.

For the noisy Poisson sensor, the sampling frequency controls the transition between these regimes. Below the kinetic matching frequency $f_0$, the observed trajectory requires positive entropy production. At matching, the minimum compatible dissipation vanishes, while the physical sensor continues to dissipate energy. The upper and lower bounds differ by less than a factor of two and establish a linear approach to zero.

The equilibrium classification is exact for the sign-symmetric four-state model, while the entropy production lower bound applies to arbitrary finite-state Markov processes with deterministic, time-reversal even binary observations. Experimentally, the bound requires sufficient temporal resolution to detect rapid visible transitions. A finite time alternative derived in the Supplemental Material [15] reduces the need for short time derivatives. Together, persistence statistics and correlations between successive dwell times provide a means to identify irreversible hidden dynamics without reconstructing the microscopic states. More broadly, these results clarify what binary records from molecular sensing, ion-channel gating, or conformational switching can certify about hidden dissipation and what they cannot identify about a particular microscopic mechanism.

# Supplemental Material

## When Time-Reversible Sensor Records Require Hidden Dissipation

Andrey V. Dobrynin*[1)] and Anastasia Stroujkova[2)]

[1)] *Department of Chemistry, University of North Carolina, Chapel Hill, NC, 27599, United States*

[2)]*SeisLogic, LLC, Chapel Hill, NC, 27517, United States*

email: avd@email.unc.edu

## S1. What a binary detector can observe

The observed trajectory contains only switches of the sensor state $Z = \pm 1$, while changes in the environmental state $Y = \pm 1$ remain unresolved. The alignment variable $U = ZY$ equals to $+1$ for aligned states and $-1$ for misaligned states. Environmental flips occur at rates $a > 0$ and $b > 0$ from aligned and misaligned states, respectively, while the corresponding sensor switching rates are $c > 0$ and $d > 0$. With rows and columns ordered as $U = +1, -1$, the matrices describing evolution at fixed $Z$ between observed switches and transitions associated with observed switches are

$$T = \begin{pmatrix} -a-c & a \\ b & -b-d \end{pmatrix}, \qquad M = \begin{pmatrix} 0 & c \\ d & 0 \end{pmatrix}, \qquad (T+M)\mathbf{1} = 0, \tag{S1.1}$$

where $\mathbf{1}$ is the column vector. Introducing

$$S = a + b + c + d, \tag{S1.2}$$

the stationary conditional probability distribution of the alignment variable for a visible state $Z$ is

$$\pi = \frac{1}{S}(b+d,\ a+c), \qquad \pi(T+M) = 0. \tag{S1.3}$$

Due to sign symmetry, the two visible states have equal stationary probabilities. The observed switching rate and additional parameter are given by

$$\mu = \pi M \mathbf{1} = \frac{ad + bc + 2cd}{S}, \qquad s = \sqrt{cd}, \tag{S1.4}$$

We consider all stationary finite-state Markov processes with transition rate matrix $Q = (q_{ij})$ and stationary probability distribution $p = (p_i)$, where $q_{ij}$ is the transition rate from state $i$ to state $j$ for $i \neq j$. Their deterministic binary observations reproduce the stationary distribution $P$ of observed trajectories, including correlations between successive dwell times. The observed variable is assumed to be even under time reversal: reversing a trajectory reverses the order of observations without changing their signs.

The minimum entropy production consistent with the observed trajectory is defined as

$$\sigma_{\inf}(P) = \inf_{Q} \sigma(Q, p), \tag{S1.5}$$

where the infimum is taken over all such processes, without requiring that it be attained. The stationary entropy production rate is given by

$$\sigma(Q,p) = k_B \sum_{i<j} (p_i q_{ij} - p_j q_{ji}) \ln \frac{p_i q_{ij}}{p_j q_{ji}}. \tag{S1.6}$$

The sum runs over all unordered pairs of connected states. A transition with a nonzero stationary flux but no reverse transition contributes infinite entropy production. Pairs with no transitions in either direction contribute zero. States with zero stationary probability are excluded.

The complete observed trajectory is characterized by the joint probability distribution of successive dwell times and switches of the binary variable. For a fixed initial visible state, the probability density of a trajectory containing $n$ observed switches separated by dwell intervals $t_0, \dots, t_n$, including possible censoring of the first and last intervals, is

$$\mathcal{L}(t_0, \dots, t_n) = \pi e^{Tt_0} M e^{Tt_1} M \cdots M e^{Tt_n} \mathbf{1}. \tag{S1.7}$$

The full stationary trajectory distribution is obtained by multiplying by the stationary probability 1/2 of the initial visible state. Unobserved Poisson refreshes, if present, do not contribute to this distribution.

For $c \neq d$, we construct the invertible observability matrix $B = (\mathbf{1}, M\mathbf{1})$:

$$B = \begin{pmatrix} 1 & c \\ 1 & d \end{pmatrix}. \tag{S1.8}$$

Direct matrix multiplication yields

$$B^{-1}TB = \begin{pmatrix} 0 & S\mu - s^2 \\ -1 & -S \end{pmatrix} \equiv X, \quad B^{-1}MB = \begin{pmatrix} 0 & s^2 \\ 1 & 0 \end{pmatrix} \equiv Y, \pi B = (1,\mu), B^{-1}\mathbf{1} = (1,0)^{\mathsf{T}}. \tag{S1.9}$$

Every path density Eq.(S1.7) is therefore

$$(1,\mu) e^{Xt_0} Y \cdots Y e^{Xt_n} (1,0)^{\mathsf{T}} \tag{S1.10}$$

and depends only on $(S, s, \mu)$. Matching these parameters reproduces the complete stationary observed trajectory, including correlations between successive dwell times. This result is consistent with the canonical forms of aggregated Markov models [6, 7, 11]. For $c = d$, the observed switching rate becomes independent of the hidden alignment, and the signal reduces to a symmetric telegraph process with $\mu = s = c$.

The converse is needed to identify all sign-symmetric four-state models that produce the same observed trajectory. The persistence

$$g(t) = \pi e^{Tt} \mathbf{1} \tag{S1.11}$$

satisfies the Cayley–Hamilton relation

$$g'' + Sg' + (S\mu - s^2) g = 0. \tag{S1.12}$$

Using$g(0) = 1$, $g'(0) = -\mu$, and $g''(0) = s^2$, we obtain

$$g'''(0) = S(\mu^2 - s^2) - \mu s^2. \tag{S1.13}$$

For $\mu \neq s$, the observed persistence uniquely determines the three parameters:

$$\mu = -g'(0), \quad s^2 = g''(0), \quad S = \frac{g'''(0) + \mu s^2}{\mu^2 - s^2}. \tag{S1.14}$$

Thus, within the sign-symmetric four-state model, identical observed trajectories necessarily have the same $S$, $s$, and $\mu$. At $\mu = s$, the persistence reduces to $g(t) = e^{-st}$, and $S$ cannot be determined from the observations. The complete dwell-time distribution also determines the observed trajectory, since

$$g'(t) = -\mu \overline{F}_\tau(t), \qquad \mu = \frac{1}{\langle \tau \rangle}, \tag{S1.15}$$

where $\tau$ is the duration of a complete dwell between consecutive observed switches and $\overline{F}_\tau(t)$ is the survival probability of a complete dwell. This property does not generally hold for other hidden-state models.

## S2. Why the recorded signal has no arrow of time

A microscopic square network can sustain a circulating probability current while the observed binary trajectory remains time reversible. We show that this holds for all stationary trajectories in the sign-symmetric four-state model. To establish this result without assuming detailed balance, we define the symmetric matrix $H$ in the canonical basis:

$$H = \begin{pmatrix} 1 & \mu \\ \mu & s^2 \end{pmatrix}. \tag{S2.1}$$

Using Eq.(S1.9), we obtain $HX = X^\mathsf{T} H$ and $HY = Y^\mathsf{T} H$, together with $(1, \mu) = \mathbf{e}_1^\mathsf{T} H$ and $\mathbf{e}_1 = (1,0)^\mathsf{T}$. Thus

$$\mathcal{L}(t_0, \ldots, t_n) = \mathbf{e}_1^\mathsf{T} H e^{X t_0} Y \cdots Y e^{X t_n} \mathbf{e}_1 = \mathcal{L}(t_n, \ldots, t_0). \tag{S2.2}$$

Since the two visible states have equal stationary probabilities, this relation establishes time reversibility of the complete observed trajectory. The matrix $H$ need not be positive definite or invertible. Thus, time reversibility of the observed trajectory does not require microscopic detailed balance. At $c = d$, the observed trajectory reduces to a symmetric telegraph process, which is also time reversible.

The determinant of the matrix $H$ is $s^2 - \mu^2$. Although the observed trajectory is time reversible for all rates, an equilibrium Markov model reproducing the same trajectory exists only for $s \geq \mu$. At $s = \mu$, the signal reduces to a symmetric telegraph process. Thus, time reversibility of the observed trajectory does not imply microscopic equilibrium.

## S3. How persistence can exclude equilibrium

The persistence $g(t)$ is the stationary probability that the observed binary state does not switch during a time interval $t$. We define

$$h = \frac{S}{2}, \qquad \omega^2 = h^2 - S\mu + s^2 = ab + \frac{(a + c - b - d)^2}{4} > 0. \tag{S3.1}$$

At equilibrium, persistence is a weighted sum of exponential decays with nonnegative weights. A negative covariance between successive persistence intervals therefore rules out equilibrium without requiring information about the hidden states.

The two eigenvalues of $T$ are $-h \pm \omega$, and direct spectral expansion gives

$$g(t) = e^{-ht}\left[\cosh(\omega t) + \frac{h-\mu}{\omega}\sinh(\omega t)\right]. \tag{S3.2}$$

Using $g(0) = 1$, $g'(0) = -\mu$, and $g''(0) = s^2$, we obtain the exact relation

$$g(2t) - g(t)^2 = \frac{s^2 - \mu^2}{\omega^2} e^{-2ht}\sinh^2(\omega t). \tag{S3.3}$$

Equations (S3.2) and (S3.3) give Eqs. (4) and (5) of the main text.

For an equilibrium Markov process, persistence is a weighted sum of exponential decays with nonnegative weights. The generator restricted to the initial visible state is self-adjoint in the stationary weighted inner product. Its survival probability $g_Q(t)$ therefore satisfies $g_Q(2t) \geq g_Q(t)^2$ for all $t \geq 0$. Equation (S3.3) violates this condition when $\mu > s$, ruling out every finite-state equilibrium Markov process with a deterministic binary observation, regardless of the number of hidden states.

At $\mu = s$, Eq.(S3.2) reduces to $g(t) = e^{-st}$. However, exponential persistence alone does not establish that the observed process is Markovian. For $c \neq d$, the stationary identity

$$\pi T = -s\pi, \qquad \pi M = s\pi. \tag{S3.4}$$

shows that both evolution without a visible switch and an observed switch preserve the same conditional stationary alignment distribution. Successive dwell times are therefore independent and exponentially distributed with rate $\mu$, and the observed trajectory reduces to a symmetric telegraph process. This corresponds to stationary (weak) lumpability [18, 19], rather than strong lumpability for arbitrary initial hidden states.

For $c = d = s$, the identities in Eq. (S3.4) need not hold. Instead, $M\mathbf{1} = s\mathbf{1}$ and $T\mathbf{1} = -s\mathbf{1}$. The observed switching rate is independent of hidden alignment, and the signal reduces to a symmetric telegraph process by strong lumpability. In both boundary cases, a four-state model with zero entropy production can be constructed by choosing $c = d = s$ and $a = b = S/2 - s > 0$. Thus, the minimum entropy production within the family of four-state models vanishes at $\mu = s$.

## S4. Building an equilibrium machine with the same record for $\mu < s$

To establish that equilibrium is possible, we construct a reversible four-state model that reproduces the complete observed trajectory. We introduce $h = S/2$ and $K = ad + bc = S\mu - 2s^2$, and define

$$k = \frac{K}{2s} > 0, \qquad q = h\sqrt{1 - \frac{\mu^2}{s^2}}, \qquad \rho = \frac{h-q}{k+s} > 0. \tag{S4.1}$$

These parameters satisfy $k + s = h\mu/s$ and $h^2 - q^2 = (k+s)^2$, giving $\rho^{-1} = (h+q)/(k+s)$ and hence $(k+s)(\rho + \rho^{-1}) = S$. The resulting positive transition rates define a second sign-symmetric four-state Markov model:

$$a_{\text{eq}} = k\rho, \quad b_{\text{eq}} = k/\rho, \quad c_{\text{eq}} = s\rho, \quad d_{\text{eq}} = s/\rho. \tag{S4.2}$$

Equations (S4.1) and (S4.2) give the equilibrium construction in Eqs. (8) and (9) of the main text.

The parameters of the new model satisfy $c_{\text{eq}}d_{\text{eq}} = s^2$, $a_{\text{eq}} + b_{\text{eq}} + c_{\text{eq}} + d_{\text{eq}} = S$, and $a_{\text{eq}}d_{\text{eq}} + b_{\text{eq}}c_{\text{eq}} = 2ks = K$. The observed switching rate $\mu$ is also unchanged. Thus, the two models produce identical stationary observed trajectories, as established in **Section S1**. For either visible state $z$, the stationary probabilities of the two hidden states are

$$p_{z,1} = \frac{1}{2(1+\rho^2)}, \qquad p_{z,2} = \frac{\rho^2}{2(1+\rho^2)}. \tag{S4.3}$$

The transition rates between hidden states satisfy $p_{z,1}a_{\text{eq}} = p_{z,2}b_{\text{eq}}$, while transitions between visible states also obey detailed balance. The constructed four-state model therefore has zero entropy production, although its joint $(Y, Z)$ dynamics may differ from those of the original model. At $\mu = s$, the symmetric telegraph process provides an equilibrium description.

## S5. Dissipation bounds from visible kinetics and finite-time persistence

### S5.1 A rate-independent bound from observable flip kinetics

We consider an arbitrary stationary finite-state Markov process with transition rate matrix $Q$ and stationary probability distribution $p$. The matrix $J$ contains transitions that change the observed binary state, while $T = Q - J$ describes evolution between observed switches, retaining the diagonal elements of $Q$. Stationarity requires $pT = -pJ$ and $T\mathbf{1} = -J\mathbf{1}$, giving

$$g(t) = pe^{Tt}\mathbf{1}, \qquad \mu = -g'(0) = pJ\mathbf{1}, \qquad s^2 = g''(0) = pJ^2\mathbf{1}. \tag{S5.1}$$

We define the forward and reverse switching rates for each microscopic state $i$:

$$c_i = \sum_j J_{ij}, \quad c_i^R = p_i^{-1}\sum_j p_j J_{ji}, \quad \sum_i p_i c_i = \mu, \quad \sum_i p_i c_i c_i^R = s^2. \tag{S5.2}$$

The stationary probability flux between states $i$ and $j$ with different visible labels is $F_{ij} = p_i J_{ij}$. Applying the Cauchy–Schwarz inequality twice gives a bound on the overlap of forward and reverse fluxes in terms of the observed persistence curvature:

$$\Omega = \sum_{i,j}\sqrt{p_i J_{ij}\, p_j J_{ji}} \le \sum_i p_i \sqrt{c_i c_i^R} \le s. \tag{S5.3}$$

The first inequality follows from applying the Cauchy–Schwarz inequality to the sum over $j$ for each state $i$, and the second from applying it with the normalized weights $p_i$.

For each unordered pair of states connected by a transition that changes the visible label, we introduce $x = F_{ij}$, $y = F_{ji}$, $w = x + y$, and $z = 2\sqrt{(xy)}/(x+y)$. The corresponding entropy production is $k_B w\Phi(z)$, where

$$\Phi(z) = 2\sqrt{1-z^2}\,\text{artanh}\sqrt{1-z^2}, \qquad 0 < z \le 1. \tag{S5.4}$$

For $v = \sqrt{1-z^2}$,

$$\Phi''(z) = \frac{2}{v^3}\left[\frac{v}{1-v^2} - \operatorname{artanh}(v)\right] > 0. \tag{S5.5}$$

Thus, $\Phi(z)$ is decreasing and convex for $0 < z \leq 1$. Since the total activity of transitions changing the observed state is $\mu$, applying Jensen's inequality together with $\Omega \leq \mathrm{s}$ gives

$$\sigma(Q,p) \geq k_B \mu \Phi\left(\frac{\Omega}{\mu}\right) \geq 2k_B \delta \operatorname{artanh}\left(\frac{\delta}{\mu}\right) = \underline{\sigma}, \tag{S5.6}$$

where we define $\delta = \sqrt{\mu^2 - s^2}$. For $\mu \leq s$, we set $\underline{\sigma} = 0$. Taking the infimum over all compatible processes gives Eq. (7) of the main text.

Transitions that do not change the observed state contribute nonnegative entropy production. Thus, $\underline{\sigma}$ provides a lower bound for any finite-state Markov model reproducing the observed trajectory, without restrictions on transition rates or symmetry. The bound depends only on the observed switching rate$\mu = -g'(0)$ and persistence curvature $s^2 = g''(0)$. It applies to unconditional stationary persistence for any deterministic binary observation, which coincides with conditional persistence in the sign-symmetric four-state model. For $\mu > s$, the bound cannot be attained by any finite-state model. **Section S7** shows that $\underline{\sigma}$ is the optimal bound based only on $\mu$ and $s$.

### S5.2 Nonattainment at finite dimension

If $\sigma(Q,p) = \underline{\sigma}$ for $\mu > s$, all transitions that preserve the visible state must have zero entropy production. Their generator $K_{\text{int}}$ therefore satisfies $pK_{\text{int}} = 0$. Using $Q = K_{\text{int}} + J - \operatorname{diag}(c_i)$ and the stationarity condition $pQ = 0$, we obtain $pJ = p \operatorname{diag}(c_i)$ , or $c_i^R = c_i$ on the stationary support. Thus,

$$s^2 = \sum_i p_i\, c_i c_i^R = \sum_i p_i\, c_i^2 \geq \left(\sum_i p_i\, c_i\right)^2 = \mu^2. \tag{S5.7}$$

This contradicts $\mu > s$. Therefore, every finite-state model reproducing the observed trajectory has $\sigma(Q,p) > \underline{\sigma}$, although $\sigma_{\text{inf}}(P) = \underline{\sigma}$ remains possible.

### S5.3 Finite-window linear certificate

For a fixed visible state $A$, we define $u(j)$ and $u^\dagger(j)$ as the probabilities of no observed switch during a forward or reversed time interval $t$, starting from microscopic state $j$. Averaging over the stationary conditional distribution $p_A$, we obtain

$$\langle u \rangle_A = \langle u^\dagger \rangle_A = g(t), \qquad \langle u u^\dagger \rangle_A = g(2t). \tag{S5.8}$$

Using $\Lambda = g^2(t) - g(2t)$, we find

$$2\Lambda = \left\langle \left(u - u^\dagger\right)^2 \right\rangle_A - \langle (\delta u)^2 \rangle_A - \left\langle \left(\delta u^\dagger\right)^2 \right\rangle_A, \tag{S5.9}$$

where $\delta u = u - \langle u \rangle_A$ and $\delta u^\dagger = u^\dagger - \langle u^\dagger \rangle_A$. Thus,

$$\Lambda(t) = g(t)^2 - g(2t) = -\mathrm{Cov}(u, u^\dagger) \le \frac{1}{4}\langle (u - u^\dagger)^2 \rangle. \tag{S5.10}$$

The forward and reversed survival probabilities correspond to the same event under $Q$ and its stationary time reversal $Q^\dagger$. Pinsker's inequality gives

$$(u - u^\dagger)^2 \le \frac{1}{2} D\left(P_j^Q \| P_j^{Q^\dagger}\right). \tag{S5.11}$$

Averaging over the stationary distribution $p$, the path relative entropy is $t\sigma(Q,p)/k_B$, since the forward and reversed processes have identical escape rates. Using equal stationary probabilities of the two visible states and their identical persistence $g(t)$, we obtain

$$\frac{1}{2}\sum_{z=\pm 1} \langle (u_z - u_z^\dagger)^2 \rangle_{p_z} \le \frac{t\sigma(Q)}{2k_B} \;\Rightarrow\; \sigma(Q) \ge \frac{8k_B}{t}\Lambda(t). \tag{S5.12}$$

$$L_{\mathrm{FT}}(P) = 8k_B \sup_{t>0} \frac{g(t)^2 - g(2t)}{t} \le \sigma_{\inf}(P). \tag{S5.13}$$

For the sensor model, the finite-time bound is weaker than $\underline{\sigma}$ but does not require short-time derivatives. It measures the probability of no observed switches, whereas the condition $Z(0) = Z(t)$ also includes trajectories with an even number of switches. Limited detector bandwidth can affect both bounds.

Unresolved rapid switches can distort the measured $\mu$ and $s^2 = g''(0)$, potentially producing a false indication of nonequilibrium. Applying the short time bound therefore requires resolving the relevant switching times. For the four-state model, the detector dead time should be much shorter than $1/\mu$ and the $1/(h + \omega)$.

## S6. Least dissipation among indistinguishable four-state models

We construct a four-state model that reproduces the complete observed trajectory with the minimum entropy production within the sign-symmetric family.

For $\mu \neq s$, **Section S1** shows that the parameters $S$, $s$, and $\mu$ uniquely characterize the observed trajectory within this family. The optimization is therefore performed over all four-state models with these parameters. The microscopic transition rates may differ from the original model, provided that the stationary trajectory of $Z$ remains unchanged.

For $\mu > s$, we define

$$\delta = \sqrt{\mu^2 - s^2}, \qquad h = S/2, \qquad K = ad + bc = S\mu - 2s^2. \tag{S6.1}$$

A four-state model reproducing the same observed trajectory is given by

$$a_0 = h - \mu + \delta, \quad b_0 = h - \mu - \delta, \quad c_0 = \mu - \delta, \quad d_0 = \mu + \delta. \tag{S6.2}$$

The constructed rates are positive since $\mu > \delta$ and $(h - \mu)^2 - \delta^2 = \omega^2 > 0$, with $h - \mu > 0$ because

$$S^2 - 2(ad + bc + 2cd) = (a + b)^2 + 2ac + 2bd + (c - d)^2 > 0. \tag{S6.3}$$

They satisfy $a_0 + c_0 = b_0 + d_0 = h$, giving equal stationary probabilities for the two alignment states. The parameters $S$, $s$, and $\mu$ are unchanged, so the constructed model reproduces the complete observed trajectory.

For the sign-symmetric four-state model, we introduce $x = |ad - bc|$ and $K = ad + bc$. The entropy production rate is

$$\sigma = \frac{k_B x}{S} \ln \frac{K + x}{K - x}, \qquad 0 \le x < K. \tag{S6.4}$$

For fixed $S$, $s$, and $\mu$, the parameter $K = S\mu - 2s^2$ remains constant. Introducing $\Delta_a = a + c - b - d$ and using $ab = \omega^2 - \Delta_a^2/4$, we obtain

$$x^2 = K^2 - 4abs^2 = S^2(\mu^2 - s^2) + s^2\Delta_a^2. \tag{S6.5}$$

At fixed $K$, Eq. (S6.4) increases monotonically with $x$. The minimum occurs at $\Delta_a = 0$, corresponding to the model Eq. (S6.2) with $x_0 = S\delta$. The minimum entropy production within the family of four-state models is therefore

$$U = \sigma_{\min}^{(4)} = k_B \delta \ln \frac{K + S\delta}{K - S\delta}, \qquad \mu > s. \tag{S6.6}$$

This establishes the minimum given in Eq. (11) of the main text.

For $\mu = s$, the observed trajectory admits a two-state equilibrium model and a four-state equilibrium model with $c = d = s$ and $a = b = S/2 - s$. For $\mu < s$, the equilibrium four-state model constructed in **Section S4** has zero entropy production. Thus, $\sigma_{\min}^{(4)} = 0$ for $\mu \le s$. When $\mu > s$, allowing additional hidden states may reduce the minimum entropy production. The general relation is

$$0 < \underline{\sigma} \le \sigma_{\inf}(P) \le U, \qquad \mu > s. \tag{S6.7}$$

This is the bound stated in Eq. (12) of the main text.

Different four-state models can produce identical observed trajectories while dissipating different amounts of energy. Thus, the observed trajectory alone does not determine the microscopic entropy production.

The minimum $U$ is exact within the family of sign-symmetric four-state models with fixed $S$, $s$, and $\mu$. For arbitrary finite-state models, Eq. (S5.6) provides a positive lower bound when $\mu > s$, while Eq. (S6.6) gives an upper bound. Whether the unrestricted infimum is attained remains unknown.

## S7. Noisy Poisson sensing

We consider a binary environment $Y = \pm 1$ switching independently at rate $\gamma$ and a sensor $Z = \pm 1$ whose state is updated at Poisson rate $f$. A newly sampled value replaces $Z$. Each update measures the environment with accuracy $(1 + \eta)/2$ relative to $Y$ $(0 < \eta < 1)$. Only changes in the sensor state are observed; updates that leave its state unchanged do not contribute to the observed trajectory. The corresponding transition rates are

$$a = b = \gamma, \quad c = \frac{f}{2}(1-\eta), \quad d = \frac{f}{2}(1+\eta). \tag{S7.1}$$

The environmental switching time is $1/\gamma$, while the average time between sensor updates is $1/f$. Varying their ratio changes whether the observed trajectory can be reproduced by an equilibrium model.

The stationary alignment correlation is $\langle YZ \rangle = f\eta/(2\gamma + f)$. The corresponding cycle affinity, probability current, and entropy production rate are

$$\mathcal{A} = 2\ln(d/c) = 4\,\mathrm{artanh}(\eta), \quad J = \frac{\gamma}{2}\langle YZ \rangle, \quad \sigma_{\mathrm{sensor}} = \frac{2k_{\mathrm{B}}\gamma f}{2\gamma + f}\eta\,\mathrm{artanh}(\eta). \tag{S7.2}$$

Substituting the transition rates from Eq.(S7.1), we obtain

$$\mu^2 - s^2 = \frac{(d-c)^2(\gamma^2 - cd)}{(2\gamma + f)^2}, \qquad f_0 = \frac{2\gamma}{\sqrt{1-\eta^2}}. \tag{S7.3}$$

Thus, $\mu > s$ for $0 < f < f_0$, $\mu = s$ at $f = f_0$, and $\mu < s$ for $f > f_0$. For $f < f_0$, we introduce $\chi = \eta\sqrt{1-(f/f_0)^2}$. Substitution into Eq. (S6.6) gives the upper bound

$$U(f) = \sigma_{\min}^{(4)}(f) = \frac{2k_{\mathrm{B}}\gamma f}{2\gamma + f}\chi\,\mathrm{artanh}(\chi), \quad 0 < f < f_0. \tag{S7.4}$$

For the original sensor model, $U(f) < \sigma_{\mathrm{sensor}}(f)$ when $f < f_0$. This gives Eq. (15) of the main text and, together with Eq. (S5.6), establishes Eq. (16)

**Table S1.** Universal lower bound $\underline{\sigma}$, finite-time lower bound $L_{\mathrm{FT}}$, four-state upper bound $U$, and actual sensor entropy production $\sigma_{\mathrm{sensor}}$ for $\eta = 1/2$. Entropy production rates are expressed in units of $k_B\gamma$.

| $f/\gamma$ | $\underline{\sigma}$ | $L_{\mathrm{FT}}$ | $U$ | $\sigma_{\mathrm{sensor}}$ |
|---|---|---|---|---|
| 1 | 0.102 | 0.023 | 0.146 | 0.183 |
| 2 | 0.036 | 0.012 | 0.064 | 0.275 |
| 2.3 | 0.001 | 0.0004 | 0.002 | 0.294 |

**Table S1** compares the lower and upper bounds on entropy production with the actual sensor entropy production for $\eta = 1/2$ at several refresh rates. For $f = \gamma$, the minimum entropy production consistent with the observed trajectory lies between $0.102k_B\gamma$ and $0.146k_B\gamma$ while the actual sensor dissipates $0.183k_B\gamma$. The finite-time lower bound is weaker but does not require short-time derivatives of the persistence. As $f$ approaches $f_0$, both lower bounds and the upper bound decrease toward zero, while the actual sensor entropy production remains positive.

In the limit $f \uparrow f_0$, we obtain

$$U(f) \sim \frac{4k_{\mathrm{B}}\gamma\eta^2}{2\gamma + f_0}(f_0 - f), \qquad \sigma_{\mathrm{sensor}}(f_0) > 0. \tag{S7.5}$$

Introducing $h = S/2$ and $\delta = \sqrt{\mu^2 - s^2}$, the minimum entropy production within the family of four-state models can be written as the sum of the universal lower bound $\underline{\sigma}$ and a positive correction:

$$U = \underline{\sigma} + 2k_B\delta \operatorname{artanh}\left(\frac{\delta}{h-\mu}\right). \tag{S7.6}$$

For the independent-telegraph sensor with $f < f_0$, the parameters satisfy $h - \mu > \mu$ and $0 < \delta < \mu$, as follows from

$$S - 4\mu = \frac{4(\gamma^2 - cd) + (c-d)^2}{S} > 0, \qquad 0 < f < f_0. \tag{S7.7}$$

Using the monotonicity of artanh, we obtain

$$U < 2\underline{\sigma} \le 2\sigma_{\inf}(P_f), \tag{S7.8}$$

where

$$\sigma_{\inf}(P_f) = \Theta(f_0 - f) \quad (f \uparrow f_0), \qquad \sigma_{\inf}(P_f) \sim k_B f\eta \operatorname{artanh}(\eta) \quad (f/\gamma \to 0). \tag{S7.9}$$

This bound (Eq. (S7.8)) applies to the independent-telegraph sensor and does not generally hold for other four-state models.

Introducing $S_0 = 2\gamma + f_0$, with fixed $\gamma > 0$ and $0 < \eta < 1$, the parameter $\delta$ near the equilibrium boundary satisfies

$$\delta^2 \sim \frac{2\eta^2\gamma^2 f_0}{S_0^2}(f_0 - f). \tag{S7.10}$$

The lower and upper bounds therefore approach zero linearly, with limiting slopes

$$\lim_{f\uparrow f_0} \frac{\underline{\sigma}(f)}{f_0 - f} = \frac{4k_B\eta^2\gamma f_0}{S_0^2}, \quad \lim_{f\uparrow f_0} \frac{U(f)}{f_0 - f} = \frac{4k_B\eta^2\gamma}{S_0}, \qquad S_0 = 2\gamma + f_0. \tag{S7.11}$$

These bounds establish that $\sigma_{\inf}(P_f)$ vanishes linearly as $f \to f_0^-$, although its exact prefactor remains unknown. In the rare-sampling limit $f/\gamma \to 0$, both bounds approach the same leading expression,

$$\sigma_{\inf}(P_f) \sim k_B f\eta \operatorname{artanh}(\eta). \tag{S7.12}$$

This is the rare-sampling asymptote given in Eq. (17) of the main text.

For the independent-telegraph sensor, the exact minimum within the family of four-state models is less than twice the unrestricted infimum. To examine the sharpness of the universal lower bound, we consider fixed $\mu > s$ and $s$ in the limit $h = S/2 \to \infty$:

$$\lim_{h\to\infty}\left[\underline{\sigma} + 2k_B\delta \operatorname{artanh}\left(\frac{\delta}{h-\mu}\right)\right] = \underline{\sigma}. \tag{S7.13}$$

The rates in Eq. (S6.2) remain positive for sufficiently large $h$ at fixed $\mu$ and $s$. Thus, $\underline{\sigma}$ is the strongest universal lower bound determined by these two parameters. However, $S$ changes along this sequence, so the models do not reproduce the same complete observed trajectory.

The entropy production bound Eq. (S5.6) also applies to the alternating renewal process of Ref. [14], since it does not require sign symmetry or a four-state model. The dwell time in state $A$ is exponentially distributed with rate $1+q$, while the dwell-time distribution in state $B$ has Laplace transform $(b_q z + c_q)/[z^2 + 2(1+q)z + c_q]$, where $b_q = 2q/(1+q)$ and $c_q = 1 + q + q^2$. The mean dwell time in state $B$ is $\langle \tau_B \rangle = 2/(1+q)$ and its probability density at the origin is $b_q$. Using the mean duration of a complete $A$–$B$ cycle, $3/(1+q)$, we obtain the unconditional stationary persistence

$$g(t) = \frac{1}{3} e^{-(1+q)t} + \frac{2}{3} g_B(t), \tag{S7.14}$$

where $g_B(t)$ is the survival probability of the remaining dwell conditional on being in state B at a stationary observation time $t$. Differentiating at $t = 0$, we obtain

$$\mu = \frac{2(1+q)}{3}, \qquad s^2 = \frac{(1+q)^2 + 2q}{3}, \qquad \mu^2 - s^2 = \frac{q^2 - 4q + 1}{9}. \tag{S7.15}$$

For $\mu > s$, we obtain $0 < q < 2 - \sqrt{3} \approx 0.268$, and Eq. (S5.6) gives a positive lower bound $V_\infty(P_q) \geq \underline{\sigma}(q) > 0$ independent of the number of hidden states and in the units of Ref. [14] ($k_B = 1$, same entropy-production convention). At $q = 1/20$, we find $\mu = 0.70$, $s^2 \approx 0.401$, $\delta \approx 0.299$, and $\underline{\sigma} \approx 0.272$. Thus, the entropy production cannot approach zero for these parameters, excluding the possibility left open in Ref. [14]. For $2 - \sqrt{3} \leq q < 1$, the bound vanishes, although Ref. [14] shows that no finite-state equilibrium model can reproduce the observed trajectory.

In the Poisson sample-and-hold model, refresh events that do not change the sensor state are unobserved. The recorded trajectory therefore contains only changes in $Z(t)$. Entropy production in Eq. (S7.2) accounts for transitions of the joint $(Y, Z)$ system but does not include additional energy costs associated with the physical sampling apparatus.

## S8. What kinetic matching means for inference and timing

A recorded sensor switch provides information about the hidden environment through both the new reading and the time elapsed since the previous switch. We denote the complete history of observed sensor switches up to time $t$ by $\mathcal{H}_t^Z$ and the average environmental state inferred from that history by $\hat{y}_t = \langle Y_t \mid \mathcal{H}_t^Z \rangle$. Here $Y_t = \pm 1$ and $Z_t = \pm 1$ are the environmental and sensor states, respectively. The inferred environmental state changes continuously while the sensor reading is held and changes abruptly when a switch is recorded [20]:

$$\frac{d\hat{y}}{dt} = -2\gamma \hat{y} + \frac{f\eta Z}{2}(1 - \hat{y}^2), \qquad \hat{y}_{\text{post}} = \frac{\hat{y}_{\text{pre}} - \eta Z_{\text{pre}}}{1 - \eta Z_{\text{pre}} \hat{y}_{\text{pre}}}. \tag{S8.1}$$

Here the subscripts "pre" and "post" refer to values immediately before and after a recorded sensor switch, respectively. The first term describes the loss of information as the environment switches independently at rate gamma. The second accounts for the information gained from the absence of an observed sensor switch. At a recorded switch, the inferred state changes according to the second expression.

To describe whether the current sensor reading agrees with the inferred environmental state, we introduce

$$x = Z\,\mathrm{artanh}(\hat{y}), \quad \beta = \mathrm{artanh}(\eta). \tag{S8.2}$$

In these variables, Eq. (S8.1) becomes

$$\frac{dx}{dt} = \frac{f\eta}{2} - \gamma \sinh(2x), \quad x_{\mathrm{post}} = \beta - x_{\mathrm{pre}} \tag{S8.3}$$

During a hold, $x$ approaches the stable value

$$x_* = \frac{1}{2}\mathrm{arsinh}\left(\frac{f\eta}{2\gamma}\right). \tag{S8.4}$$

Each recorded switch replaces x by $\beta - x$. The two processes restrict the stationary values of $x$ to the interval between $x_*$ and $\beta - x_*$. To see that these are the actual limits, a long hold brings $x$ arbitrarily close to $x_*$, a subsequent switch brings it close to $\beta - x_*$, and the evolution during the next hold passes through all intermediate values. The switch rate $f[1 - \eta\tanh(x)]/2$ stays positive for $0 < \eta < 1$.

At $f = f_0$, $x_* = \beta/2$ and the interval reduces to one value. The environmental state inferred from the complete sensor history then depends only on the present reading:

$$\hat{y}_t = Z_t \tanh\left(\frac{\beta}{2}\right), \tag{S8.5}$$

The stationary sensor output is a symmetric telegraph process with switching rate gamma. This is stationary (weak) lumpability; arbitrary initial distributions of the hidden states need not produce a Markov sensor trajectory. Importantly, the sensor does not become perfectly accurate at matching: the inferred environmental state still changes sign when $Z$ changes.

The interval also limits the confidence of any observer who sees only the sensor trajectory. The largest attainable magnitude of the inferred environmental state is

$$H(f) = \tanh(\max\{x_*, \beta - x_*\}), \tag{S8.6}$$

This limit is smallest at matching, where $H(f_0) = \tanh(\beta/2)$. At other refresh rates, the timing of sensor switches can provide additional information about the environment. Over a prediction time tau, the independent environmental switching reduces the maximum confidence to $\exp(-2\gamma\tau)\,H(f)$.

A downstream memory that receives only the recorded sensor trajectory cannot exceed this limit. For example, suppose a decision requires an inferred environmental state at time $t + \tau$ with magnitude greater than $\theta$. No sensor history-based decision can meet this requirement if

$$\exp(-2\gamma\tau)\,H(f) \le \theta. \tag{S8.7}$$

The range of refresh rates, for which the decision is excluded, can be obtained explicitly. Introducing $u = \theta\exp(2\gamma\tau)$ and $\ell = \mathrm{artanh}(u)$, for

$$\tanh(\beta/2) < u < \eta, \tag{S8.8}$$

the condition holds throughout $f_- \le f \le f_+$, where

$$f_- = \frac{2\gamma}{\eta}\sinh[2(\beta - \ell)], f_+ = \frac{2\gamma}{\eta}\sinh(2\ell). \quad \text{(S8.9)}$$

These limits concern prediction from the observed sensor history. They do not assume that a physical downstream memory reproduces the optimal estimate, and they do not apply if an additional observation clock or an independent measurement is available.

The same switching dynamics can be tested directly from observed dwell times. A complete dwell is the time $\tau$ between two consecutive changes of $Z$. Since each dwell begins immediately after a switch, its initial hidden-state distribution differs from the distribution at an arbitrary stationary time. Using the no-switch matrix $T$, the observed-switch matrix $M$, and the stationary conditional alignment distribution $\pi$ introduced in **Section S1**, we define

$$\alpha = \frac{\pi M}{\pi M \mathbf{1}}, \qquad R = (-T)^{-1} \quad \text{(S8.10)}$$

Here $\alpha$ is the probability distribution of the alignment variable $U = YZ$ immediately after an observed sensor switch. The probability that a complete dwell lasts longer than t is

$$\Pr(\tau > t) = \alpha \exp(Tt)\, \mathbf{1}. \quad \text{(S8.11)}$$

Its first two moments and the correlation between successive complete dwell times follow as

$$\langle \tau \rangle = \alpha R \mathbf{1}, \qquad \langle \tau^2 \rangle = 2\alpha R^2 \mathbf{1}, \qquad \mathrm{Cov}(\tau_i, \tau_{i+1}) = \alpha R^2 M R^2 M \mathbf{1} - \langle \tau \rangle^2 \quad \text{(S8.12)}$$

The normalized covariance used in Fig. 2 is $\mathrm{Cov}(\tau_i, \tau_{i+1})/\langle \tau \rangle^2$. For the independent environment ($a = b = \gamma$), the excess dwell-time dispersion $\mathrm{CV}^2 - 1$ has the sign of $cd - \gamma^2$, while the covariance between adjacent dwells has the opposite sign. Both vanish at $f = f_0$. At matching, complete dwell times are independent exponentials. These complete-dwell measurements must not be confused with stationary persistence $g(t)$, which begins at a randomly chosen time rather than just after a recorded switch.

A second test uses sensor autocorrelation. With $z_i = \pm 1$ for the visible label in microscopic state $i$, and $Q$ and $p$ the full four-state generator and stationary distribution, respectively, we obtain

$$C_Z(t) = \sum_{i,j} p_i\, z_i [\exp(Qt)]_{ij} z_j, \;\; \Delta(t) = C_Z(2t) - C_Z(t)^2. \quad \text{(S8.13)}$$

For an equilibrium Markov process with a time-reversal-even observation, $\Delta(t)$ cannot be negative. In the sample-and-hold sensor it is negative for every finite $t > 0$ when $f < f_0$, excluding an equilibrium explanation of the observed trajectory.

The correlation can be written explicitly. For $f \neq 2\gamma$,

$$C_Z(t) = A\exp(-ft) + (1 - A)\exp(-2\gamma t), \qquad A = \frac{f^2(1 - \eta^2) - 4\gamma^2}{f^2 - 4\gamma^2}. \quad \text{(S8.14)}$$

At $f = 2\gamma$ the two decay rates coincide, but the correlation remains finite:

$$C_Z(t) = (1 + \gamma\eta^2 t)\exp(-2\gamma t). \quad \text{(S8.15)}$$

Correlations between more distant dwells provide another test.

For $a = b = \gamma$, introducing

$$D_0 = \gamma(c + d) + cd, \qquad \lambda = -\frac{cd}{D_0}, \tag{S8.16}$$

we obtain

$$\mathrm{Cov}(\tau_n, \tau_{n+m}) = \mathrm{Cov}(\tau_n, \tau_{n+1})\lambda^{m-1}, \qquad m \geq 1. \tag{S8.17}$$

Because $-1 < \lambda < 0$, correlations between progressively separated dwells alternate in sign and decrease in magnitude. This prediction extends the adjacent-dwell comparison in Fig. 2 of the main text.

The confidence bound above assumes symmetric sensor readings. For an asymmetric sensor, the information gained from successive switches changes, and the interval and matching expressions need not hold. This restriction concerns prediction calculation, not the general equilibrium criterion for four-state models derived in the preceding sections.

## S9. Figure 2 of the main text

Figure 2 of the main text tests several logically distinct predictions with one stochastic trajectory model: the spectral obstruction, the dwell statistics, and actual microscopic dissipation. The numerical markers were generated from exact continuous-time Gillespie trajectories of the four-state $(Y, Z)$ process, without fitting the parameters or rescaling Monte Carlo observables to analytical predictions. At $\gamma = 1$, $\eta = 1/2$, eight sampling rates were used: $f/\gamma = 0.6, 1, 1.5, 2, f_0/\gamma, 3, 4, 5$. Each rate was simulated for 15 independent stationary observation windows of duration $T = 3 \times 10^5/\gamma$, after a burn-in period of $10^3/\gamma$. The spectral witness was measured from $Z$ sampled at spacing $0.05/\gamma$ with $\gamma t = 0.5$. The entropy production was calculated from microscopic jump log-rate ratios, and complete dwell dispersion and serial covariance were evaluated from successive observed sensor-flip times.

Error bars in Fig. 2 of the main text are 95% confidence intervals based on the 15 independent trajectory estimates, not independent individual time samples. The exact curves follow the generators and formulas above. The equilibrium emulator can additionally be simulated independently for $f > f_0$; its path-law equality is established in **Sections S1–S4**.

The simulations used MATLAB with the Mersenne Twister generator (seed 20261022), 15 independent stationary observation windows per rate, and a stored per-replicate output structure. The analysis script and saved replicate estimates are provided in the accompanying numerical archive.

In Fig. 2(b) of the main text, the green curve is $\underline{\sigma}$ from Eq. (S5.6), the blue dashed curve is U from Eq. (S6.6), and black with circles denotes the actual sensor. The curves $\underline{\sigma}$ and U are exact analytical bounds, not fits.

The numerical examples are consistency checks of independently simulated records against exact calculations. They do not by themselves prove either all-dimensional nonrealizability or equality of every possible path probability; those statements follow from the preceding analytical arguments.